\documentclass{article}

 \usepackage[preprint]{neurips_2026}

\usepackage[utf8]{inputenc} 
\usepackage[T1]{fontenc}    
\usepackage{hyperref}       
\usepackage{url}            
\usepackage{booktabs}       
\usepackage{amsfonts}       
\usepackage{nicefrac}       
\usepackage{microtype}      
\usepackage{xcolor}         

\usepackage[textsize=tiny]{todonotes}

\usepackage{amsmath}
\usepackage{amssymb}
\usepackage{mathtools}
\usepackage{amsthm}

\theoremstyle{plain}
\newtheorem{theorem}{Theorem}[section]

\theoremstyle{definition}
\newtheorem{definition}[theorem]{Definition}

\theoremstyle{remark}

\usepackage{siunitx} 
\usepackage{pifont} 
\usepackage{xspace} 
\usepackage{algorithm}
\usepackage{algorithmic}
\usepackage{float}
\usepackage{adjustbox} 
\usepackage{titlesec}
\usepackage{etoolbox}
\usepackage{xcolor} 
\usepackage{pgfplots}
\pgfplotsset{compat=1.18}
\usepackage{tikz}
\usetikzlibrary{arrows.meta,automata,positioning}
\usepackage{pgf-pie}
\usepackage{pgf-pie}
\usepackage[most]{tcolorbox}

\newcommand{\frameworkname}{\textsc{JustDiag}\xspace}

\newcommand{\faultinsight}{\textit{FaultInsight}\xspace}
\newcommand{\rcagent}{\textit{RCAgent}\xspace}
\newcommand{\flowofaction}{\textit{Flow-of-Action}\xspace}

\newcommand{\styledstd}[1]{{\scriptsize\textcolor{gray}{$\pm$#1}}}

\definecolor{djgreen}{HTML}{1B5E20}
\definecolor{djblue}{HTML}{1E3A8A}
\definecolor{djyellow}{HTML}{8A5A00}
\definecolor{djred}{HTML}{8B1E1E}
\definecolor{djpurple}{HTML}{6A1B9A}
\definecolor{djteal}{HTML}{0F766E}
\definecolor{djgray}{HTML}{4B5563}

\newcommand{\djtt}[2]{\text{\texttt{\textcolor{#1}{#2}}}}
\newcommand{\supportsrel}{\djtt{djgreen}{supports}}
\newcommand{\contradictsrel}{\djtt{djred}{contradicts}}
\newcommand{\insufficientrel}{\djtt{djyellow}{insufficient}}
\newcommand{\resolvedstatus}{\djtt{djgreen}{resolved}}
\newcommand{\provresolvedstatus}{\djtt{djteal}{provisionally\_resolved}}
\newcommand{\needmorestatus}{\djtt{djyellow}{need\_more\_evidence}}
\newcommand{\stalledstatus}{\djtt{djblue}{stalled}}

\newcommand{\alternativetorel}{\djtt{djpurple}{alternative\_to}}
\newcommand{\selectedoverrel}{\djtt{djblue}{selected\_over}}

\usepackage{caption}
\DeclareCaptionFont{algofont}{\footnotesize} 
\titlespacing*{\section}{0pt}{0.6ex plus 0.2ex minus 0.1ex}{0.4ex}
\titlespacing*{\subsection}{0pt}{0.5ex plus 0.2ex minus 0.1ex}{0.3ex}
\everydisplay\expandafter{\the\everydisplay\small}
\AtBeginEnvironment{equation}{\small}
\AtBeginEnvironment{equation*}{\small}
\AtBeginEnvironment{align}{\small}
\AtBeginEnvironment{align*}{\small}
\AtBeginEnvironment{gather}{\small}
\AtBeginEnvironment{gather*}{\small}
\AtBeginEnvironment{multline}{\small}
\AtBeginEnvironment{multline*}{\small}

\tcbset{
  artifactbox/.style={
    enhanced,
    colback=blue!4!white,
    colframe=blue!75!black,
    coltitle=white,
    fonttitle=\bfseries,
    boxrule=0.9pt,
    arc=2.5mm,
    left=7pt,right=7pt,top=6pt,bottom=6pt
  },
  promptbox/.style={
    enhanced,
    colback=green!3!white,
    colframe=green!45!black,
    coltitle=white,
    fonttitle=\bfseries,
    boxrule=0.9pt,
    arc=2.5mm,
    left=7pt,right=7pt,top=6pt,bottom=6pt
  },
  statbox/.style={
    enhanced,
    colback=orange!4!white,
    colframe=orange!70!black,
    coltitle=white,
    fonttitle=\bfseries,
    boxrule=0.9pt,
    arc=2.5mm,
    left=7pt,right=7pt,top=6pt,bottom=6pt
  },
  eventbox/.style={
    enhanced,
    colback=purple!4!white,
    colframe=purple!60!black,
    coltitle=white,
    fonttitle=\bfseries,
    boxrule=0.9pt,
    arc=2.5mm,
    left=7pt,right=7pt,top=6pt,bottom=6pt
  }
}

\title{JustDiag!: A Diagnostic Justification Engine for Accountable Root Cause Analysis}

\author{%
Tingzhu Bi \\
Peking University \\
\texttt{bitingzhu@stu.pku.edu.cn} \\
\And
Xinrui Jiang \\
Peking University \\
\texttt{jxrjxrjxr@pku.edu.cn} \\
\And
Xun Zhang \\
Peking University \\
\texttt{zhangxun@stu.pku.edu.cn} \\
\AND
Pengcheng Su \\
Peking University \\
\texttt{pcs@pku.edu.cn} \\
\And
Congjie He \\
University of Edinburgh \\
\texttt{congjie.he@ed.ac.uk} \\
\And
Jinglin Li \\
Beijing University of \\ Posts and Telecommunications \\
\texttt{jlli@bupt.edu.cn} \\
\AND
Ping Wang \\
Peking University \\
\texttt{pwang@pku.edu.cn} \\
\And
Meng Ma \\
Peking University \\
\texttt{mameng@pku.edu.cn} \\
}

\begin{document}

\maketitle

\begin{abstract}
    Large language models can produce fluent root cause analyses, but fluent final answers alone are insufficient evidence for accountability in high-stakes operations. In real incident response, engineers need to know what evidence supported a diagnosis, which alternatives were considered, where contradictions remained, and whether the system resolved the case or preserved uncertainty. We address this gap with \frameworkname, a diagnostic justification engine for RCA that maintains an explicit process state over evidence, findings, competing hypotheses, conflicts, and next checks. We evaluated the system on 66 real-world incidents using a two-layer protocol that separately scores final-answer quality and process quality. Relative to a matched control without diagnostic justification, \frameworkname achieved stronger outcome and process scores, while accepting slightly lower terminal completion due to more calibrated non-closure. These results suggest that accountable RCA requires explicit diagnostic justification artifacts and process-aware evaluation, not only fluent final answers. 
\end{abstract}
  
  \section{Introduction}   
\label{sec:introduction}

Root cause analysis (RCA) in large-scale operational systems is a high-stakes task because operators must decide not only what likely failed, but also whether the available evidence justifies action. Modern incidents emerge from interactions among compute, storage, network, and application subsystems whose dependencies are only partially visible in telemetry. As a result, practical RCA rarely reduces to reading off a single anomalous metric; it requires comparing explanations, challenging weak hypotheses, and sometimes deferring closure when uncertainty remains.

This makes accountability central rather than optional. Recent LLM-based RCA systems produce fluent reports and useful investigations~\cite{wang2024rcagent,pei2025flow,roy2024exploring,wang2025towards,zhang2025thinkfl}, but most evaluations still focus on the final answer. That view is insufficient in high-stakes settings, where operators must also know what evidence was used, which alternatives were considered, how contradictions were handled, and whether the system preserved or hid residual uncertainty.

We address this gap with \emph{diagnostic justification}: an explicit process artifact over evidence, findings, competing hypotheses, claims, conflicts, and next checks. We instantiate this idea in \frameworkname, a diagnostic justification engine for RCA in which an orchestrator coordinates domain experts, decomposes hypotheses into verification claims, adjudicates those claims as \supportsrel, \contradictsrel, or \insufficientrel, and exports a structured diagnosis state rather than only a final report. Because accountable behavior is not fully visible in the final answer alone, we evaluate both \emph{outcome-only} quality and \emph{process-aware} quality using silver reference narratives and structured judge inputs.

On 66 real-world incidents, \frameworkname improved Outcome Score from 51.0 to 57.7 and Process Score from 44.0 to 50.5 relative to a matched no-DJ control, while accepting a modest reduction in terminal completion from 65/66 to 62/66. This suggests that calibrated non-closure can be preferable to premature resolution when the available evidence does not justify a clean diagnosis.

Taken together, these choices reposition RCA from answer generation toward accountable adjudication. The central question is no longer only whether the system can produce a plausible diagnosis, but whether it can expose enough structured evidence for operators to trust, contest, or defer that diagnosis responsibly.

Our contributions are threefold:
\begin{enumerate}
    \item We introduce \textbf{diagnostic justification} as a framing for accountable RCA and instantiate it in a multi-agent system that maintains explicit evidence, findings, competing hypotheses, conflicts, and next checks.
    \item We develop a \textbf{claim-level adjudication framework} for comparing competing hypotheses under uncertainty, including termination states such as \resolvedstatus, \provresolvedstatus, \stalledstatus, and \needmorestatus.
    \item We provide a \textbf{process-aware evaluation protocol and empirical study} on 66 real-world incidents showing that diagnostic justification improves both outcome quality and process quality relative to a matched no-DJ control.
\end{enumerate}

The remainder of the paper proceeds as follows. Section~\ref{sec:related_work} situates our work relative to RCA systems, process-aware evaluation, and accountability-oriented assessment. Section~\ref{sec:method} presents the diagnostic-justification framework and its adjudication mechanics. Section~\ref{sec:evaluation} describes the evaluation protocol and reports the main, ablation, and sensitivity results.

\section{Related Work}
\label{sec:related_work}

\paragraph{RCA and LLM-based diagnostic systems.}
Traditional AIOps methods largely cast RCA as fault localization, dependency analysis, or causal ranking. Representative systems such as \textit{RCD}~\cite{ikram2022root} and \textit{$\epsilon$-diagnosis}~\cite{shan2019diagnosis}, together with related causal or statistical approaches~\cite{li2022causal,pan2021faster}, help identify anomalous components and likely bottlenecks from telemetry. Tools such as \faultinsight~\cite{bi2024faultinsight} further improve observability structure and can narrow the search space before deeper investigation. In parallel, recent LLM-based RCA systems have moved toward narrative synthesis and tool-augmented investigation, including single-agent approaches such as \rcagent~\cite{wang2024rcagent}, broader agentic RCA formulations~\cite{roy-et-al-2024-llmagents-rca,roy2024exploring,wang2025towards}, and SOP-style or multi-agent systems such as \flowofaction~\cite{pei2025flow} and \textit{mABC}~\cite{zhang2024mabc,zhang2025thinkfl}. These lines improve either localization or flexibility, but most still end at a final diagnosis or ranked candidate set rather than exporting an auditable artifact that records evidence use, alternative handling, and unresolved uncertainty.

\paragraph{Process-aware evaluation and reasoning artifacts.}
A growing literature shows that plausible final answers do not guarantee trustworthy reasoning traces. In language models, chain-of-thought and related rationale artifacts can be post-hoc, incomplete, or only weakly faithful to the true basis of a prediction~\cite{lanham-et-al-2023-faithfulness,turpin2023language}. This observation has motivated broader interest in structured search and process supervision, including \textit{Tree of Thoughts}~\cite{yao2023tree}, process-supervision approaches~\cite{lightman2023let}, and rationale-oriented benchmarks such as \textit{ERASER}~\cite{deyoung-et-al-2020-eraser}. Across these directions, the common shift is to ask whether intermediate reasoning artifacts contain information worth supervising, auditing, or evaluating directly rather than treating them as auxiliary text. Our work follows this process-aware perspective, but with a different target: we do not evaluate generic free-form chain-of-thought. Instead, we focus on operationally meaningful structures such as evidence links, competing hypotheses, contradiction records, and next checks, because these are the parts of the process that matter most for accountable RCA.

\paragraph{Accountability, oversight, and high-stakes deployment.}
Work on interpretability, accountability, and high-stakes AI argues that trustworthy deployment requires more than predictive performance. Prior work emphasizes the need for explanations that support interrogation and contestation~\cite{miller2019explanation,doshi-velez-kim-2017-rigorous}, as well as reviewable records, traceable decision pathways, and governance mechanisms that make system behavior auditable~\cite{kroll-et-al-2017-accountable,raji-et-al-2020-closinggap,cobbe-lee-singh-2021-reviewable,kroll-2021-traceability,singh-cobbe-norval-2019-decisionprovenance}. Similar expectations appear in broader risk-management and reporting frameworks for high-stakes AI~\cite{nist-2023-airmf,openai-2023-gpt4systemcard}. These arguments are especially relevant to RCA, where outputs may trigger rollback, throttling, traffic shifting, or service restart under uncertainty. We extend this perspective by treating diagnostic artifacts themselves as accountability evidence: not explanations after the fact, but structured records that help operators decide whether a diagnosis is ready to trust, ready to contest, or not yet ready to close.

\section{Method: Diagnostic Justification for Accountable RCA}
\label{sec:method}

We formulate root cause analysis (RCA) as the construction of an explicit \emph{diagnostic justification} rather than the generation of a final answer alone. In high-stakes operations, a plausible diagnosis is insufficient if operators cannot inspect what evidence supported it, which alternatives were considered, where contradictions remained, and whether the system terminated because the case was resolved or because uncertainty was still unresolved. Our method therefore maintains a structured diagnostic state throughout the investigation and treats RCA as an adjudication process over competing explanations.

At a high level, the system consists of an orchestrator that coordinates domain-specific experts and maintains a shared justification state. Experts retrieve and interpret telemetry evidence, derive intermediate findings, and return structured assessments about verification claims. The orchestrator aggregates these updates, revises the current-best hypothesis and plausible alternatives, records conflicts and unresolved uncertainty, and determines whether the case should be marked as \resolvedstatus, \provresolvedstatus, \needmorestatus, or \stalledstatus. The final output is therefore not only a diagnosis report, but also an auditable process artifact summarizing why the system reached that state.

\begin{figure*}[t]
  \centering
  \includegraphics[width=\textwidth]{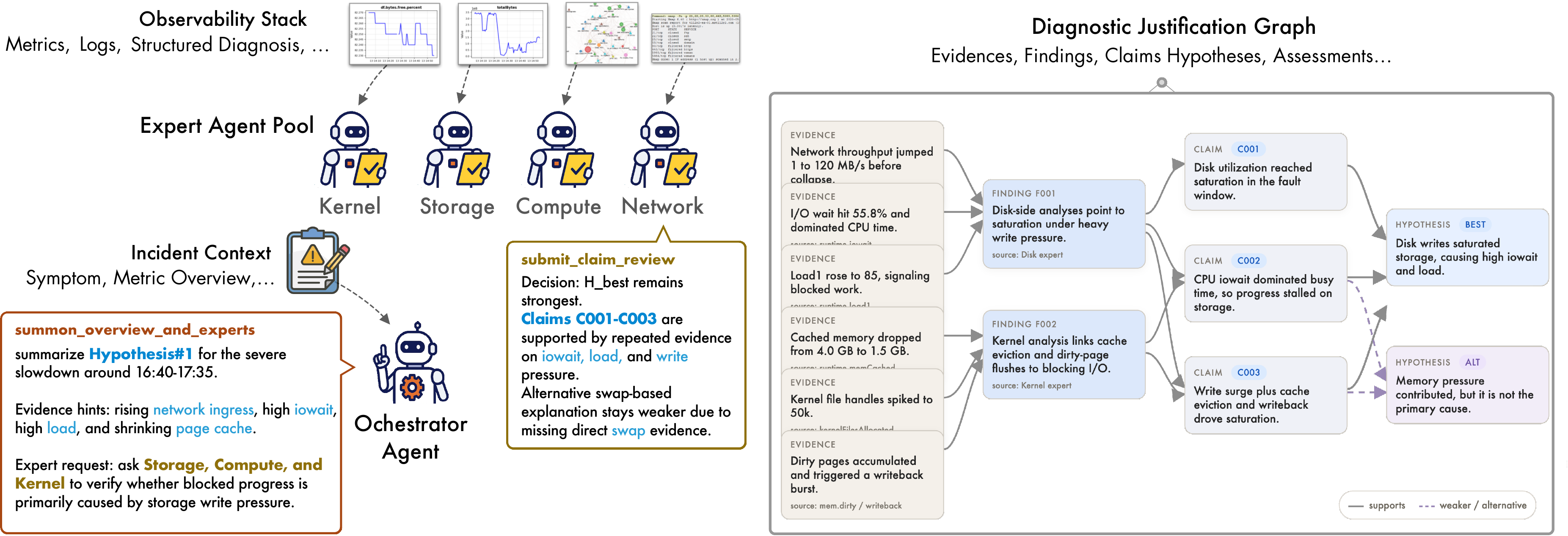}
  \caption{\textbf{\frameworkname as a diagnostic justification engine.}
  The orchestrator combines incident context with observability summaries, dispatches domain experts, and maintains a shared diagnostic justification state. Expert outputs are converted into evidence, findings, claims, and assessments, which are then integrated into a diagnostic justification graph for adjudication over competing hypotheses.}
  \label{fig:overview-justdiag}
\end{figure*}

Figure~\ref{fig:overview-justdiag} illustrates this workflow. The left side shows how the orchestrator grounds its coordination decisions in incident context and observability summaries before dispatching experts. The right side shows the resulting diagnostic justification graph, where evidence, findings, claims, hypotheses, and assessments are maintained as explicit reviewable objects rather than being left implicit in dialogue. This design supports reviewability in two senses: it helps operators inspect the reasoning process during diagnosis, and it enables later evaluation of process quality independently from final-answer quality.

\paragraph{Primitive Components}
We define five primitive objects that together form the diagnostic justification artifact. 

\begin{definition}
\label{def:evidence}
An \emph{evidence} item $e$ is a grounded observation extracted from the observability space $\mathcal{D}$. We represent it as
\[
e = \langle id, source, span, desc \rangle,
\]
where $source$ identifies the underlying telemetry or tool output, $span$ localizes the relevant region within that source, and $desc$ is a natural-language description of the observed phenomenon.
\end{definition}

\begin{definition}
\label{def:finding}
A \emph{finding} $f$ is an intermediate diagnostic statement derived from one or more evidence items. We represent it as
\[
f = \langle id, \mathcal{E}_f, \alpha, \rho, content \rangle,
\]
where $\mathcal{E}_f \subseteq \mathcal{E}$ is the supporting evidence set, $\alpha$ is the producing expert or agent, $\rho \in [0,1]$ is a confidence score, and $content$ is the finding statement.
\end{definition}

\begin{definition}
\label{def:hypothesis}
A \emph{hypothesis} $h$ is a candidate causal explanation for the incident. We represent it as
\[
h = \langle id, ctx, content, status \rangle,
\]
where $ctx$ records the diagnostic context in which the hypothesis was introduced, $content$ is the explanatory statement, and $status$ denotes its current adjudication state, such as current-best, plausible-alternative, rejected, or underspecified.
\end{definition}

\begin{definition}
\label{def:claim}
A \emph{claim} $c$ is a discriminative substatement attached to a hypothesis and intended for verification against available evidence. We represent it as
\[
c = \langle id, h, priority, content \rangle,
\]
where $h$ is the parent hypothesis, $priority$ indicates how important the claim is for adjudication, and $content$ states what must be checked.
\end{definition}

\begin{definition}
\label{def:assessment}
An \emph{assessment} $a$ is a structured evaluation of a claim. We represent it as
\[
a = \langle c, v, \gamma, p, r \rangle,
\]
where $c$ is the target claim, $v \in \{\supportsrel, \contradictsrel, \insufficientrel\}$ is the verdict, $\gamma \in [0,1]$ is a confidence score, $p$ is a priority label, and $r$ is a natural-language rationale.
\end{definition}

\subsection{Diagnostic Justification Graph}
\label{sec:dj_graph}

At step $t$, the RCA process is represented by a diagnostic justification state
\begin{equation}
\mathcal{S}_t = (\mathcal{E}_t, \mathcal{F}_t, \mathcal{H}_t, \mathcal{C}_t, \mathcal{A}_t, \mathcal{U}_t),
\end{equation}
where $\mathcal{E}_t$ is the set of evidence items observed so far, $\mathcal{F}_t$ is the set of derived findings, $\mathcal{H}_t$ is the set of active hypotheses, $\mathcal{C}_t$ is the set of verification claims attached to those hypotheses, $\mathcal{A}_t$ is the set of structured claim assessments returned by experts, and $\mathcal{U}_t$ stores unresolved diagnostic state, including conflicts, missing distinguishing evidence, remaining uncertainties, and recommended next checks.

We externalize this state as a typed directed graph
\begin{equation}
G_t = (V_t, E_t, \lambda_V, \lambda_E),
\end{equation}
where
\[
V_t = \mathcal{E}_t \cup \mathcal{F}_t \cup \mathcal{H}_t \cup \mathcal{D}_t.
\]
Here $\mathcal{E}_t$, $\mathcal{F}_t$, $\mathcal{H}_t$, and $\mathcal{D}_t$ contain evidence, finding, hypothesis, and decision nodes, respectively. The map $\lambda_V$ assigns node types and optional node states, while $\lambda_E$ assigns edge semantics such as \supportsrel, \contradictsrel, \alternativetorel, and \selectedoverrel. The resulting graph is not a proof object. Instead, it is a reviewable process artifact that records support, contradiction, competition, and decision state as the diagnosis unfolds.

\paragraph{Relational Schema}
The justification artifact is organized around coupled support and adjudication relations rather than a single monotonic derivation tree. In the graph, evidence supports findings and findings support or contradict hypotheses:
\[
(e,f)\in E_t \Rightarrow \lambda_E(e,f)=\supportsrel, \qquad
(f,h)\in E_t \Rightarrow \lambda_E(f,h)\in\{\supportsrel,\contradictsrel\}.
\]
This yields the operational chain $e \rightarrow f \rightarrow h$ while preserving explicit contradiction links.

The adjudication structure is layered on top of this chain. Each hypothesis $h$ carries a set of attached claims $\mathcal{C}_t(h) \subseteq \mathcal{C}_t$, where claims serve as structured, testable substatements rather than additional evidence nodes. Each assessment targets exactly one claim, giving a relation $a \rightsquigarrow c$ between expert judgments and the discriminative units they evaluate. Terminal labels are then produced by adjudicating the resulting hypothesis state, so decision nodes summarize why a hypothesis was preferred rather than acting as logical children in a derivation tree.
\begin{figure}[t]
  \centering
  \begin{tikzpicture}[
    >=Latex,
    box/.style={draw, rounded corners, minimum height=0.95cm, minimum width=1.95cm, align=center, font=\small},
    lbl/.style={font=\scriptsize, inner sep=1pt}
  ]
  \node[box] (e) {Evidence\\$e$};
  \node[box, right=2.5cm of e] (f) {Finding\\$f$};
  \node[box, right=3.1cm of f] (h) {Hypothesis\\$h$};
  \draw[->, thick] (e) -- node[above, lbl] {\supportsrel} (f);
  \draw[->, thick] (f) -- node[above, lbl] {\supportsrel{} / \contradictsrel} (h);

  \node[box, below=0.95cm of h, xshift=-1.25cm] (c) {Claim\\$c\in\mathcal{C}(h)$};
  \node[box, left=2.2cm of c] (a) {Assessment\\$a$};
  \draw[->, thick, dashed] (h) -- node[left, lbl, pos=0.35] {decompose} (c);
  \draw[->, thick] (a) -- node[above, lbl] {verdict on} (c);

  \node[box, below=0.95cm of h, xshift=2.25cm, minimum width=2.8cm] (tau) {Terminal status\\$\tau(\mathcal{S})$};
  \draw[->, thick] (h) -- node[right, lbl, pos=0.55] {adjudicate} (tau);
  \end{tikzpicture}
  \caption{\textbf{Diagnostic justification relations.}
  Evidence supports findings; findings support or contradict hypotheses; claims decompose hypotheses; experts emit structured assessments of individual claims; the orchestrator adjudicates competing hypotheses and assigns terminal status.}
  \label{fig:dj-relational-schema}
\end{figure}
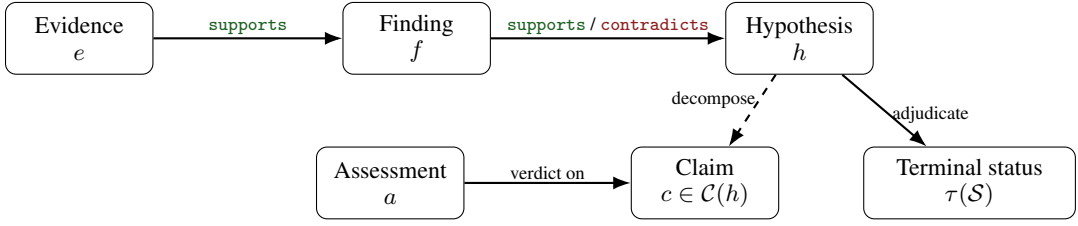

\subsection{State Update and Graph Evolution}
\label{sec:state_update}

The justification artifact evolves in discrete rounds. Let $\Delta_t$ denote the structured increment produced at round $t$:
\[
\Delta_t =
\bigl(\Delta^{\mathrm{add}}_t,\;
      \Delta^{\mathrm{link}}_t,\;
      \Delta^{\mathrm{adj}}_t,\;
      \Delta^{\mathrm{hyp}}_t,\;
      \Delta^{\mathrm{evt}}_t\bigr),
\]
where $\Delta^{\mathrm{add}}_t$ introduces new evidence, findings, hypotheses, or decisions; $\Delta^{\mathrm{link}}_t$ introduces or revises typed edges; $\Delta^{\mathrm{adj}}_t$ appends claim assessments; $\Delta^{\mathrm{hyp}}_t$ updates hypothesis metadata such as ranks and statuses; and $\Delta^{\mathrm{evt}}_t$ appends an audit event.

We write the coupled update as
\begin{equation}
(G_{t+1},\mathcal{S}_{t+1}) = \textsc{UpdateStateAndGraph}(G_t,\mathcal{S}_t,\Delta_t),
\end{equation}
where the update operator adds new nodes and edges, appends claim assessments, and refreshes hypothesis metadata such as ranks and statuses. If the round introduces dangling references or stale duplicated support/contradiction edges, they are immediately cleaned up so that the exported graph remains internally consistent. 

Claim-level adjudication is folded into this update. For each claim $c$ under hypothesis $h$, the orchestrator accumulates assessments into an adjudication vector
\begin{equation}
A(c,h) = \big(\mathrm{Sup}(c,h),\mathrm{Con}(c,h),\mathrm{Ins}(c,h)\big),
\end{equation}
and ranks hypotheses by
\begin{equation}
\mathrm{Score}(h) = \sum_{c \in \mathcal{C}_t(h)} \bigl(\mathrm{Sup}(c,h) - \mathrm{Con}(c,h)\bigr).
\end{equation}
The purpose of this score is not to estimate a calibrated posterior probability. It is to maintain a reviewable preference ordering over competing explanations while preserving unresolved contradiction and insufficiency signals.
 
\subsection{Termination Semantics}
\label{sec:termination}

We define a termination function
\begin{equation}
\tau(\mathcal{S}_t) \in \{\resolvedstatus, \provresolvedstatus, \needmorestatus, \stalledstatus\},
\end{equation}
where the output depends on the current-best hypothesis, claim coverage, unresolved conflicts, missing distinguishing evidence, and whether further discriminative checks remain available. We define the current-best hypothesis as
\[
h_t^\star = \arg\max_{h \in \mathcal{H}_t} \mathrm{Score}(h),
\]
and define claim coverage as the fraction of claims under $h$ that have received at least one structured assessment:
\[
\mathrm{Cov}(h) =
\frac{\left|\{c \in \mathcal{C}_t(h) : \exists a \in \mathcal{A}_t \text{ with target } c\}\right|}{|\mathcal{C}_t(h)|}.
\]
Let $\mathrm{Conf}(h_t^\star)$ denote the key unresolved conflicts for $h_t^\star$, and let $\mathrm{Miss}(h_t^\star)$ denote its missing distinguishing evidence.

The system is \resolvedstatus{} when $h_t^\star$ is strongly supported, $\mathrm{Cov}(h_t^\star)$ is adequate, $\mathrm{Conf}(h_t^\star)$ is negligible, and $\mathrm{Miss}(h_t^\star)=\varnothing$. It is \provresolvedstatus{} when one explanation is clearly favored but minor ambiguity remains. It is \needmorestatus{} when closure is not yet justified but further discriminative checks remain actionable. It is \stalledstatus{} when the system cannot responsibly claim clean resolution and further discriminative progress appears limited, for example because the remaining conflicts are persistent or the available next checks have largely been exhausted.

These terminal states separate diagnostic quality from diagnostic certainty. A stalled case is therefore not a null output: it still exports a current-best hypothesis, plausible alternatives, key conflicts, remaining uncertainties, and recommended next checks.

\subsection{Overall Procedure}
\label{sec:overall_procedure}  

Algorithm~\ref{alg:dj_adjudication} summarizes the overall procedure. The orchestrator initializes candidate hypotheses from incident context and early telemetry, decomposes them into claims, dispatches expert investigations, updates the diagnostic justification graph and state with newly produced evidence, findings, and assessments, re-ranks hypotheses, and evaluates the termination function on the updated state.

\noindent
\begin{minipage}[t]{0.40\linewidth}
\vspace{0pt}
Read at a high level, the loop alternates between evidence collection and adjudication. Each round selects the current-best explanation, dispatches experts to gather targeted evidence, and merges their findings and claim assessments into the shared justification state. The orchestrator then refreshes the live hypothesis set and re-evaluates termination, so closure is always a consequence of updated support, contradiction, and coverage rather than a purely conversational stopping rule.

\vspace{0.5em}
This side-by-side layout is intentional: the algorithm is meant as a compact procedural summary of the state-and-graph dynamics already defined above, not as a second source of semantics. Readers can therefore use it as an operational guide while following the surrounding prose for the meaning of each object and transition.
\end{minipage}\hfill
\begin{minipage}[t]{0.56\linewidth}
\vspace{-5pt}
\begin{algorithm}[H]
\footnotesize
\caption{Diagnostic Justification via Graph and State Updates}
\label{alg:dj_adjudication}
\textbf{Input:} Incident context $\mathcal{I}$, observability data $\mathcal{D}$, expert pool $\Pi$, budget $B$. \\
\textbf{Output:} Final report $R$, justification state $\mathcal{S}$, and justification graph $G$.
\begin{algorithmic}[1]
\STATE Initialize justification state $\mathcal{S}_0$ and graph $G_0$
\STATE $\mathcal{H}_0 \leftarrow \textsc{InitialTriage}(\mathcal{I}, \mathcal{D})$
\STATE $\mathcal{C}_0 \leftarrow \textsc{GenerateClaims}(\mathcal{H}_0)$
\STATE $(G, \mathcal{S}) \leftarrow (G_0, \mathcal{S}_0)$
\STATE $t \leftarrow 0$
\STATE $\tau \leftarrow \needmorestatus$
\WHILE{$t < B$ \textbf{and} $\tau = \needmorestatus$}
    \STATE $h^\star \leftarrow \textsc{SelectCurrentBest}(\mathcal{S})$
    \STATE $\Pi_t \leftarrow \textsc{SelectExperts}(h^\star, \mathcal{S}, \Pi)$
    \STATE $\Delta_t \leftarrow \emptyset$
    \FOR{each expert $\pi \in \Pi_t$}
        \STATE $(E_\pi, F_\pi, A_\pi) \leftarrow \pi.\textsc{Investigate}(\mathcal{D}, h^\star, \mathcal{S})$
        \STATE $\Delta_t \leftarrow \Delta_t \cup E_\pi \cup F_\pi \cup A_\pi$
    \ENDFOR
    \STATE $(G, \mathcal{S}) \leftarrow \textsc{UpdateStateAndGraph}(G, \mathcal{S}, \Delta_t)$
    \STATE $\mathcal{S} \leftarrow \textsc{RefreshHypothesesIfNeeded}(\mathcal{S})$
    \STATE $\tau \leftarrow \textsc{EvaluateTermination}(\mathcal{S})$
    \STATE $t \leftarrow t + 1$
\ENDWHILE
\STATE $R \leftarrow \textsc{GenerateReport}(\mathcal{S}, \tau)$
\STATE \textbf{return} $(R, \mathcal{S}, G)$
\end{algorithmic}
\end{algorithm}
\end{minipage}

  \section{Experimental Evaluation}
  \label{sec:evaluation}

  We evaluated \frameworkname on a real-world industrial incident dataset using a protocol that separated final-answer quality from justification quality. The experiments addressed four questions. \textbf{RQ1} asked whether diagnostic justification improved outcome quality and process quality relative to a matched no-DJ control and two external agent baselines. \textbf{RQ2} asked which DJ mechanisms mattered most by ablating claim adjudication, evidence grounding, hypothesis competition, and the coverage gate. \textbf{RQ3} asked how sensitive the method was to key design choices, including claim count, hypothesis count, and coverage threshold. \textbf{RQ4} asked whether calibrated non-closure yielded a more accountable diagnosis state when the evidence did not yet justify a clean resolution.
  
  \subsection{Experimental Setup}
  
  \begin{figure*}[t]
      \centering
      \includegraphics[width=\linewidth]{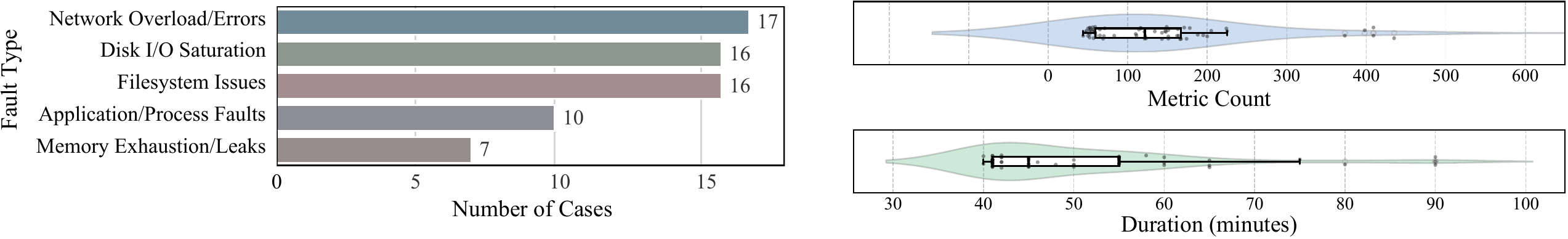}
      \caption{Dataset Overview. (Left) Distribution of fault types shows the diversity of the dataset. (Right) Violin plots illustrate the distribution of key metrics across all incidents.}

      \label{fig:dataset_overview:a}
      \label{fig:dataset_overview:b}
      \label{fig:dataset_overview}
  \end{figure*}

  \paragraph{Dataset} We used a proprietary dataset of 66 real-world incidents from a hyperscale data center. Each case included multivariate metric telemetry and an SRE-written silver reference narrative. As shown in Figure~\ref{fig:dataset_overview}, the dataset covered diverse failure modes and high-dimensional observability patterns.
  
  \vspace{-0.3em}
  \paragraph{Implementation Details \& Reproducibility}
  \vspace{-0.3em}
  Unless otherwise specified, we used Gemini 3 Flash as the diagnosis backbone and GPT-5.4 as the evaluation judge. The main comparison in RQ1 used the full 66-case dataset. The ablation and sensitivity studies in RQ2--RQ3 used a 10-case mini subset for cost control. We therefore treated the mini-set results as directional mechanism checks rather than directly comparable replacements for the full main results. We provide a minimal code-and-data package with the core \frameworkname{} MVP, a runnable \texttt{case7.yaml} example, the corresponding \texttt{case7.parquet} sample file, and exported case artifacts; detailed artifact examples, prompt excerpts, and extended tables are included in the appendix.

  \paragraph{Evaluation Metrics}
  We reported two judge-based metrics and a set of run-level statistics. \emph{Outcome Score} evaluated the final diagnosis artifact against the silver reference narrative. \emph{Process Score} evaluated the exported diagnostic justification artifact for evidence grounding, alternative handling, contradiction management, traceability, and uncertainty exposure. We also reported termination profile, runtime, turn count, tool calls, and token usage. In the tables below, \texttt{Comp.}, \texttt{Res.}, \texttt{Prov.}, and \texttt{Stall} denote completed, resolved, provisionally resolved, and stalled cases, respectively; \texttt{FindLink}, \texttt{Hyp. Cnt}, \texttt{Hyp. Cov}, and \texttt{Con} denote the finding-to-evidence link ratio, average active hypothesis count, hypothesis coverage ratio, and aggregate contradiction mass; and \texttt{Calls} denotes tool calls. For the full-system comparisons, \texttt{completed} referred to runs that reached an adjudication-style terminal artifact; this notion was directly comparable for \frameworkname and its matched no-DJ control, but less meaningful for external baselines that did not share the same terminal schema.
  
  \paragraph{Baselines}
  We compared \frameworkname against a matched control without diagnostic justification (\texttt{Main w/o DJ}) and two external agent baselines, \rcagent~\cite{wang2024rcagent} and \flowofaction~\cite{pei2025flow}. The primary matched comparison in the paper was \frameworkname versus \texttt{Main w/o DJ}; the external baselines served primarily as reference points for whether a justification-engine design improved over existing agentic RCA styles.

  \begin{table*}[t]
      \centering
      \caption{\textbf{Main and mechanism results.} Panel A reports the full 66-case main comparison used for RQ1. Panel B reports 10-case mini-subset ablations used for RQ2. Because Panels A and B use different datasets, they should not be compared numerically across panels; the mini-set results are directional mechanism checks. In Panel A, \texttt{-} denotes not applicable for external baselines that do not export adjudication-style terminal states.}
      \label{tab:main_ablation_results}
      \scriptsize
      \setlength{\tabcolsep}{3pt}
      \renewcommand{\arraystretch}{0.96}
      \adjustbox{width=\textwidth,center}{%
      \begin{tabular}{@{}lcccccccccc@{}}
        \toprule
        & \multicolumn{2}{c}{\textbf{Judge-Based Quality}} & \multicolumn{4}{c}{\textbf{Termination Profile}} & \multicolumn{4}{c}{\textbf{Cost / Effort}} \\
        \cmidrule(lr){2-3} \cmidrule(lr){4-7} \cmidrule(lr){8-11}
        \textbf{Method} & \textbf{Outcome} & \textbf{Process} & \textbf{Comp.} & \textbf{Res.} & \textbf{Prov.} & \textbf{Stall} & \textbf{Time (s)} & \textbf{Turns} & \textbf{Calls} & \textbf{Tokens} \\
        \midrule
        \multicolumn{11}{@{}l}{\textit{Panel A: Main results on the 66-case full dataset}} \\
        \textbf{JustDiag (ours)} & \textbf{57.7} & \textbf{50.5} & 62/66 & 61 & 1 & 3 & 457.0 & 6.5 & 33.7 & 218k \\
        Main without DJ control & 51.0 & 44.0 & \textbf{65/66} & \textbf{65} & 0 & 0 & 315.7 & 5.8 & 23.5 & 154k \\
        RCAgent & 44.3 & 9.5 & - & - & - & - & \textbf{64.3} & \textbf{2.7} & \textbf{6.6} & \textbf{22k} \\
        Flow-of-Action & 42.8 & 9.3 & - & - & - & - & 276.3 & 4.0 & 19.7 & 93k \\
        \midrule
        \multicolumn{11}{@{}l}{\textit{Panel B: Ablations on the 10-case mini subset}} \\
        & \multicolumn{2}{c}{\textbf{Judge-Based Quality}} & \multicolumn{2}{c}{\textbf{Termination}} & \multicolumn{4}{c}{\textbf{DJ Structure Signals}} & \multicolumn{2}{c}{\textbf{Effort}} \\
        \cmidrule(lr){2-3} \cmidrule(lr){4-5} \cmidrule(lr){6-9} \cmidrule(lr){10-11}
        \textbf{Variant} & \textbf{Outcome} & \textbf{Process} & \textbf{Comp.} & \textbf{Stall} & \textbf{FindLink} & \textbf{Hyp. Cnt} & \textbf{Hyp. Cov} & \textbf{Con} & \textbf{Turns} & \textbf{Calls} \\
        \textbf{JustDiag (mini)} & \textbf{53.6} & \textbf{53.8} & 8/10 & 2 & 0.800 & 3.1 & 0.408 & 9.7 & 6.9 & 33.8 \\
        Without claim adjudication & 45.9 & 45.6 & 9/10 & 1 & 0.782 & 1.0 & \textbf{1.000} & 0.0 & 6.7 & 25.5 \\
        Without evidence grounding & 49.1 & 35.4 & \textbf{10/10} & 0 & 0.000 & \textbf{3.7} & 0.000 & 3.3 & 5.9 & \textbf{40.2} \\
        Without hypothesis competition & \textbf{55.8} & 50.4 & \textbf{10/10} & 0 & \textbf{0.830} & 1.1 & \textbf{1.000} & 0.0 & 5.4 & 24.7 \\
        Without coverage gate & 54.1 & 50.1 & 9/10 & 1 & 0.772 & 3.1 & 0.350 & 5.9 & \textbf{4.9} & 21.3 \\
        \bottomrule
      \end{tabular}%
      }
    \end{table*}
  
  \subsection{RQ1: Main Results}
  
  Table~\ref{tab:main_ablation_results} (Panel A) supports two claims. First, the strongest causal comparison is \frameworkname versus the matched no-DJ control. Relative to \texttt{Main without DJ control}, the full system improved Outcome Score from 51.0 to 57.7 and Process Score from 44.0 to 50.5, while reducing terminal completion only slightly from 65/66 to 62/66. This pattern is consistent with the design goal of calibrated non-closure: the DJ-enabled system improved both final-answer quality and justification quality, but became more willing to preserve unresolved uncertainty when clean closure was not well supported. Second, the no-DJ control still outperformed the two external baselines, which suggests that the underlying agent architecture was already competitive. The DJ mechanisms therefore acted as an additional improvement layer rather than as the sole source of performance. The external baselines are shown with \texttt{-} in the terminal-state columns because they do not export the same adjudication-oriented terminal schema.
  
  \subsection{RQ2: Mechanism Study}
  
  Table~\ref{tab:main_ablation_results} (Panel B) reports the mechanism ablations on the 10-case mini subset. The clearest accountability-critical mechanism was evidence grounding. Removing it dropped Process Score from 53.8 to 35.4 and collapsed the finding-link signal to 0.000, even though the Outcome Score decreased only moderately to 49.1. This suggests that plausible narratives can still be produced without explicit grounding, but the exported process artifact becomes much less auditable. Claim adjudication was the second most important mechanism. Removing it reduced Outcome Score from 53.6 to 45.9 and Process Score from 53.8 to 45.6, while collapsing hypothesis plurality from 3.1 to 1.0 and eliminating contradiction mass entirely. By contrast, removing hypothesis competition or the coverage gate slightly improved outcome on this mini subset (55.8 and 54.1, respectively) but weakened the structured DJ story by reducing explicit competition, contradiction handling, or caution around closure. We therefore interpret evidence grounding and claim adjudication as the main process-forming mechanisms, while competition and coverage gating primarily function as safeguards against premature closure.

  \begin{table*}[t]
      \centering
      \caption{\textbf{Sensitivity analysis on the 10-case mini subset.} Panel A varies the number of claims per hypothesis, Panel B varies the number of candidate hypotheses, and Panel C varies the coverage threshold $\tau$. These settings were used for RQ3 and were interpreted as robustness checks over key DJ design choices. Here \texttt{Out.} and \texttt{Proc.} denote Outcome Score and Process Score, respectively, and \texttt{Calls} denotes tool calls.}
      \label{tab:sensitivity_results}
      \scriptsize
      \setlength{\tabcolsep}{3pt}
      \renewcommand{\arraystretch}{0.96}
      \begin{minipage}[t]{0.32\textwidth}
      \centering
      \captionof*{table}{\textbf{Panel A.} Claim count $C$}
      \vspace{-0.35em}
      \begin{adjustbox}{max width=\linewidth,center}
      \begin{tabular}{@{}lcccccc@{}}
        \toprule
        \textbf{C} & \textbf{Out.} & \textbf{Proc.} & \textbf{Time} & \textbf{Turns} & \textbf{Calls} \\
        \midrule
        2 & 52.0 & 43.2 & 344.3 & 5.4 & 21.9 \\
        3 & \textbf{60.8} & 51.2 & 424.9 & 5.9 & 27.8 \\
        4 & \textbf{60.8} & \textbf{60.6} & 693.0 & \textbf{8.5} & \textbf{39.9} \\
        \bottomrule
      \end{tabular}
      \end{adjustbox}
      \end{minipage}\hfill
      \begin{minipage}[t]{0.32\textwidth}
      \centering
      \captionof*{table}{\textbf{Panel B.} Hypothesis count $K$}
      \vspace{-0.35em}
      \begin{adjustbox}{max width=\linewidth,center}
      \begin{tabular}{@{}lcccccc@{}}
        \toprule
        \textbf{K} & \textbf{Out.} & \textbf{Proc.} & \textbf{Time} & \textbf{Turns} & \textbf{Calls} \\
        \midrule
        2 & \textbf{63.6} & 53.1 & 449.0 & 7.7 & 32.1 \\
        3 & 55.7 & 53.9 & 513.5 & 8.3 & 37.9 \\
        4 & 53.2 & \textbf{60.0} & \textbf{1213.4} & \textbf{11.3} & \textbf{56.1} \\
        2--4 & 59.8 & 54.1 & 615.5 & 6.9 & 36.8 \\
        \bottomrule
      \end{tabular}
      \end{adjustbox}
      \end{minipage}\hfill
      \begin{minipage}[t]{0.32\textwidth}
      \centering
      \captionof*{table}{\textbf{Panel C.} Coverage threshold $\tau$}
      \vspace{-0.35em}
      \begin{adjustbox}{max width=\linewidth,center}
      \begin{tabular}{@{}lcccccc@{}}
        \toprule
        \textbf{$\tau$} & \textbf{Out.} & \textbf{Proc.} & \textbf{Time} & \textbf{Turns} & \textbf{Calls} \\
        \midrule
        0.5 & 59.7 & \textbf{55.4} & 421.1 & \textbf{7.3} & 29.0 \\
        0.67 & 59.5 & 48.2 & 411.2 & 6.4 & 27.6 \\
        0.8 & \textbf{63.9} & 48.4 & \textbf{641.6} & 6.7 & \textbf{33.1} \\
        \bottomrule
      \end{tabular}
      \end{adjustbox}
      \end{minipage}
  \end{table*}

  \subsection{RQ3: Sensitivity to DJ Design Choices}
  
  Table~\ref{tab:sensitivity_results} shows that the DJ hyperparameters controlled interpretable quality--cost trade-offs rather than behaving arbitrarily. For claim count, $C=2$ was clearly too weak, whereas $C=3$ offered the best efficiency-quality trade-off and $C=4$ produced the strongest Process Score (60.6) at substantially higher cost. For hypothesis count, $K=2$ achieved the best Outcome Score (63.6), while $K=4$ achieved the best Process Score (60.0) but required by far the largest runtime and tool budget. The adaptive 2--4 setting was a reasonable compromise but did not dominate either extreme. For the coverage threshold, looser coverage ($\tau=0.5$) produced the best Process Score (55.4), whereas stricter coverage ($\tau=0.8$) produced the best Outcome Score (63.9). We interpret this difference as evidence that stricter closure criteria may help final-answer alignment on the mini subset, while looser criteria may preserve more explicit uncertainty and alternative handling in the justification artifact.

  \subsection{RQ4: Calibrated Non-Closure and Case Study}

RQ4 asks whether calibrated non-closure yields a more accountable diagnosis state when the evidence does not yet justify a clean conclusion. The preceding results suggest that diagnostic justification does not merely improve average scores; it also changes \emph{how} the system behaves under unresolved ambiguity. In particular, the full system was slightly less likely to terminate in a clean completed state than the matched no-DJ control (62/66 vs.\ 65/66), yet still achieved higher Outcome and Process Scores. We interpret this pattern as evidence of calibrated non-closure rather than simple failure: when discriminative evidence remains incomplete, the DJ-enabled system is more willing to preserve uncertainty explicitly instead of collapsing prematurely to a single narrative. This is a first-class part of the paper's headline claim: better RCA is not only about producing a stronger final answer, but also about exposing when clean closure is not yet justified.

\paragraph{Case study: why a stalled diagnosis can still be useful.}
Case~7 illustrates this distinction. The incident exhibited a strong connection surge together with severe system load and downstream storage symptoms, making it easy for a conventional pipeline to collapse early to a storage-only explanation. Indeed, the matched no-DJ control produced a \resolvedstatus{} diagnosis centered on disk saturation and I/O wait. By contrast, \frameworkname terminated as \stalledstatus, but exported a substantially richer justification artifact: a current-best hypothesis focused on application-level saturation under a connection flood, two plausible alternatives, explicit finding-to-hypothesis conflicts, missing distinguishing evidence, and concrete recommended next checks. In other words, the system did not return an empty or indecisive output; it returned a structured record of \emph{what was currently favored, what remained plausible, and what evidence would be needed to close the case responsibly}.

\begin{figure*}[t]
    \centering
    \includegraphics[width=\textwidth]{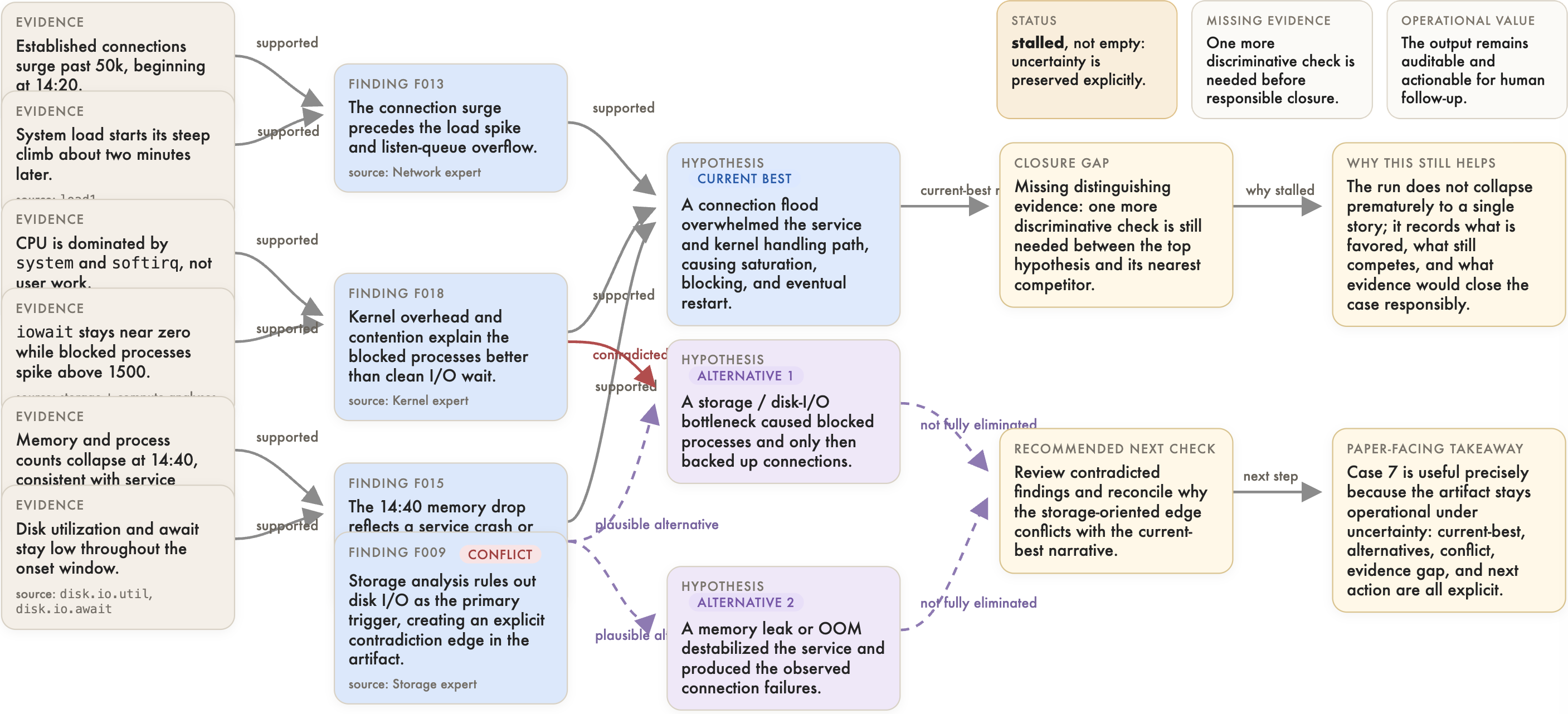}
    \caption{\textbf{Case~7: calibrated non-closure as an explicit diagnostic artifact.}
    The graph summarizes the evidence--finding--hypothesis chain exported by \frameworkname for a stalled case. The current-best explanation attributes the failure to application-level saturation under a connection flood, while storage and memory-centered alternatives remain visible rather than being silently discarded. The figure also makes the closure gap explicit: one distinguishing check is still needed to eliminate the nearest competitor responsibly. This is the key paper-facing point of the case study: the system preserves a reviewable diagnosis state even when it does not claim clean resolution.}
    \label{fig:case7-calibrated-nonclosure}
\end{figure*}

Figure~\ref{fig:case7-calibrated-nonclosure} shows the corresponding artifact. Several pieces of evidence support a network-triggered saturation narrative, and one conflict explicitly weakens the storage-first explanation. At the same time, the alternative hypotheses are not fully eliminated, because the remaining evidence is still insufficient to rule them out cleanly. The system therefore marks the case as stalled, records the missing distinguishing evidence, and recommends a concrete next check instead of forcing a single resolved story. In this sense, the case study is not merely illustrative; it operationalizes the paper's main claim that accountability sometimes requires preserving a useful diagnosis state rather than overstating certainty.

This behavior is valuable because Case~7 is precisely the kind of incident in which premature closure can be misleading. The full system identified that some storage symptoms were real but still insufficient to eliminate competing explanations about application saturation and kernel-level contention. Rather than hiding this ambiguity behind a confident single-cause report, it preserved the unresolved structure directly in the exported artifact. This is the practical merit of diagnostic justification: even when the system does not claim a clean resolution, it can still produce an auditable and operationally useful diagnosis state for human follow-up. We provide the full diagnostic graph and justification trace for this case in the appendix.

\section{Conclusion}
\vspace{-0.3em}

We present \frameworkname, a diagnostic justification engine for accountable RCA built around explicit evidence, alternatives, contradictions, and stopping criteria rather than final-answer generation alone. Across 66 real-world incidents, the method improved both outcome quality and process quality relative to a matched no-DJ control, while also exposing a meaningful quality--accountability trade-off through calibrated non-closure. These results suggest that high-stakes RCA benefits not only from stronger answers, but from explicit diagnostic artifacts that help operators inspect, contest, and responsibly act on model outputs.

\section{Limitations}
\vspace{-0.3em}

\frameworkname remains limited by the observability it receives and by the quality of the silver reference narratives used for evaluation. It may still miss latent business or deployment causes that are not visible in telemetry, and the reported judge scores should therefore be interpreted as structured proxies rather than ground truth.

\section{Broader Impacts}
\vspace{-0.3em}

This work may improve accountability in AI-assisted incident diagnosis by exposing evidence, alternatives, and uncertainty instead of only returning a final answer. However, such systems should support operators rather than replace them, because over-trust in automated RCA could still propagate incorrect diagnoses in high-stakes production settings.

\bibliographystyle{plain}
\bibliography{references}


\appendix
\newpage
\section{Diagnostic Justification Artifact Schema}
\label{sec:appendix-schema}

The central object in our method is not a hidden chain-of-thought trace, but an explicit diagnostic justification artifact. This artifact records what evidence was collected, how findings were formed, which hypotheses were compared, where contradictions emerged, and why the final state was resolved or left unresolved. The same artifact also provides the structured input to our process-aware judge.

\subsection{Core Objects}

\begin{table}[h]
\centering
\caption{Core components of the diagnostic justification artifact.}
\label{tab:appendix-artifact-schema}
\begin{tabular}{p{0.20\linewidth}p{0.72\linewidth}}
\toprule
\textbf{Component} & \textbf{Role in the artifact} \\
\midrule
Evidence & Raw metric observations or externally supplied facts that anchor subsequent findings. \\
Findings & Interpreted local statements extracted from evidence, typically describing anomalies, correlations, or candidate mechanisms. \\
Hypotheses & Candidate root-cause explanations that compete for support during adjudication. \\
Conflicts & Localized contradictions between findings and hypotheses, preserved rather than silently erased. \\
Adjudication state & Claim-level support, contradiction, and insufficiency statistics used to rank competing hypotheses. \\
Decision state & Final or provisional termination outcome, including unresolved uncertainty and recommended next checks. \\
\bottomrule
\end{tabular}
\end{table}

These objects are serialized into exportable files rather than left implicit in a raw conversation trace. This design choice matters for two reasons. First, it makes the system auditable by human reviewers. Second, it allows process-aware evaluation to operate on stable, method-agnostic artifacts instead of brittle prompt transcripts.

\subsection{What the Exported Artifact Retains}

At minimum, the exported justification state preserves the following fields:

\begin{itemize}
    \item the current best hypothesis and its textual content;
    \item plausible alternatives that remain live at termination;
    \item key supporting findings and key conflicts;
    \item missing distinguishing evidence and remaining uncertainties;
    \item recommended next checks for unresolved or provisionally resolved cases;
    \item claim-level adjudication summaries over support, contradiction, and insufficiency.
\end{itemize}

Together, these fields make it possible to distinguish between a cleanly supported diagnosis and a cautious but still unresolved one. This distinction is central to our accountability framing, because a high-stakes RCA system should be able to preserve uncertainty without collapsing into either silence or unjustified certainty.

\newpage
\section{Method Lineage and Design Rationale}
\label{sec:appendix-lineage}

\subsection{Why We Moved Away from Proof-Completion Framing}

Our earlier framing emphasized proof construction and derivation completion. In practice, this framing was too rigid for real operational incidents. Production failures rarely present a single clean logical chain from observation to root cause. Instead, they often contain overlapping symptoms, partially compatible causal stories, and unresolved ambiguity even after substantial investigation. A strict proof-completion view therefore risks rewarding brittle closure behavior over calibrated diagnosis.

We replaced that framing with diagnostic justification. The new framing asks a simpler and more operational question: what structured evidence, findings, competing hypotheses, conflicts, and uncertainty signals are necessary to make an RCA output accountable? This change better matches how incident reviews are actually conducted in practice, where engineers compare explanations, discard weak ones, preserve unresolved alternatives, and decide whether further checks are required before acting.

\subsection{Design Principles}

The current system is organized around four design principles:

\begin{enumerate}
    \item \textbf{Evidence before narrative.} Final explanations should be grounded in explicit evidence and findings rather than written directly as free-form stories.
    \item \textbf{Competition before commitment.} Candidate hypotheses should compete under claim-level adjudication rather than being accepted in first-write fashion.
    \item \textbf{Conflicts should be preserved.} Contradictory findings and unresolved tensions are part of accountable reasoning and should remain visible in the exported artifact.
    \item \textbf{Uncertainty is a first-class outcome.} Unresolved states such as \texttt{stalled} or \texttt{need\_more\_evidence} are legitimate outputs when discriminative support is insufficient.
\end{enumerate}

These principles motivate the specific mechanisms evaluated in our ablation study: evidence grounding, claim adjudication, hypothesis competition, and coverage-based termination control.

\subsection{From Principles to Mechanisms}

The implementation maps these principles into concrete system behavior. Evidence grounding links findings to observed metrics. Claim adjudication aggregates support, contradiction, and insufficiency at the claim level. Hypothesis competition prevents immediate lock-in to a single explanation. Termination semantics separate clean resolution from provisional or unresolved outcomes. Finally, exportable justification artifacts make all of this legible to both judges and human auditors.

This decomposition also clarifies the role of our control experiment, \texttt{Main w/o DJ}. The base architecture remains relatively strong even without full DJ control, which is why the main causal comparison in the paper is not against only external baselines. Instead, \texttt{Main w/o DJ} isolates the incremental value of explicit justification mechanisms within the same overall framework.

\newpage
\section{Artifact Examples and Mechanism Demonstrations}
\label{sec:appendix-cases}

This section complements the main paper by showing what \frameworkname actually exports at the case level. Rather than only presenting aggregate scores, we include concrete examples of ground-truth narratives, delivered diagnosis artifacts, paper-facing prompt excerpts, and graph-evolution signals. Our goal is to make the appendix read like an inspection pack for the diagnostic justification engine.

\subsection{Case Portfolio}

\begin{table}[h]
\centering
\caption{Appendix case portfolio for artifact-level inspection.}
\label{tab:appendix-case-portfolio}
\begin{tabular}{p{0.18\linewidth}p{0.16\linewidth}p{0.56\linewidth}}
\toprule
\textbf{Case} & \textbf{Outcome type} & \textbf{Why it is useful} \\
\midrule
\texttt{case7} & stalled & Calibrated non-closure with explicit alternatives, conflict retention, and one unresolved distinguishing check. \\
\texttt{case6\_2} & resolved & Compact resolved artifact with hypothesis competition, claim review, and a short but complete graph export. \\
\bottomrule
\end{tabular}
\end{table}

\subsection{Case 7: Ground Truth vs. Exported Artifact}

\begin{tcolorbox}[artifactbox,title={Case 7: silver reference narrative vs.\ exported diagnosis}]
\begin{minipage}[t]{0.48\linewidth}
\textbf{Silver reference narrative.}\\
\small
A sudden connection storm overwhelmed the host, causing severe load, blocked processes, and secondary storage symptoms. The incident should be read as a network-triggered saturation event rather than as a purely storage-origin failure.
\end{minipage}\hfill
\begin{minipage}[t]{0.48\linewidth}
\textbf{\frameworkname exported artifact.}\\
\small
Termination status: \texttt{stalled}.\\
Current best: application-level saturation under a connection flood.\\
Retained alternatives: storage bottleneck; memory-leak / OOM explanation.\\
Key gap: one additional discriminative check is still needed before clean closure.
\end{minipage}
\end{tcolorbox}

\begin{tcolorbox}[statbox,title={Case 7: appendix-facing stats summary}]
\small
\textbf{Session time:} 748.8s \quad
\textbf{Turns:} 11 \quad
\textbf{Tool calls:} 71 \quad
\textbf{LLM calls:} 52 \quad
\textbf{Tokens:} 281,653 \\
\textbf{Viewed metrics:} 49 / 121 (\(40.5\%\)) \quad
\textbf{Nodes:} 139 \quad
\textbf{Edges:} 164 \quad
\textbf{Debug signals:} 113 \\
\textbf{Alternatives retained:} 2 \quad
\textbf{Key conflicts:} 1 \quad
\textbf{Coverage ratio (best hypothesis):} 1.0
\end{tcolorbox}

\begin{tcolorbox}[eventbox,title={Case 7: graph-evolution / event excerpt}]
\small
\textbf{Event 1: alternative payload proposed.} The orchestrator explicitly registered two competing hypotheses: a storage bottleneck explanation and a memory-leak / OOM explanation. The recorded reasoning excerpt states that the core question was the direction of causality: did connections cause the load spike, or did an underlying bottleneck cause connections to pile up?\\[0.35em]
\textbf{Event 2: expert claim assessment.} During expert verification, the network expert returned a \texttt{supports} assessment for the connection-storm hypothesis while simultaneously contradicting storage-oriented claims about blocked processes being primarily explained by I/O wait.\\[0.35em]
\textbf{Event 3: latest termination assessment.} The final debug state marked the case as \texttt{stalled}, retained one explicit key conflict (\texttt{finding:F009}), and recorded one missing distinguishing evidence item: an additional discriminative check between the top hypothesis and the nearest competitor.
\end{tcolorbox}

\begin{figure}[p]
    \centering
    \includegraphics[width=\textwidth,height=0.92\textheight,keepaspectratio]{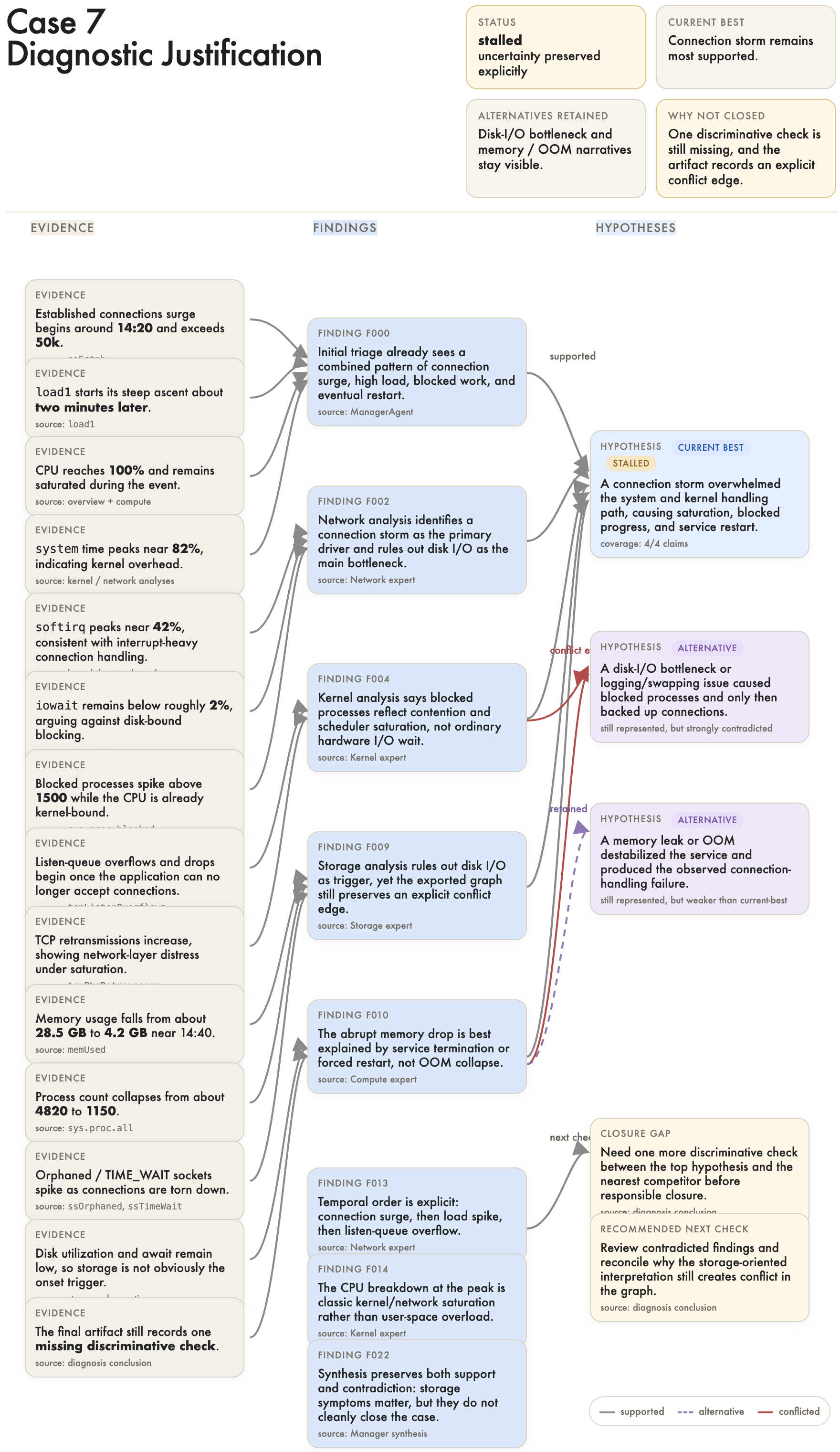}
    \caption{\textbf{Case 7 full diagnostic justification artifact.}
    This appendix-scale view exposes the evidence column, synthesized findings, current-best hypothesis, retained alternatives, closure gap, and next-check recommendation in one place. Relative to the main-text crop, the full-page version makes clear that the artifact remains operational even under \texttt{stalled} termination: it still records what was most supported, what was contradicted, and what evidence would be needed for responsible closure.}
    \label{fig:appendix-case7-a4}
\end{figure}

\subsection{Case 6\_2: Ground Truth vs. Exported Artifact}

\begin{tcolorbox}[artifactbox,title={Case 6\_2: silver reference narrative vs.\ exported diagnosis}]
\begin{minipage}[t]{0.48\linewidth}
\textbf{Silver reference narrative.}\\
\small
A burst of inbound traffic and concurrent sessions exhausted host resources, drove CPU saturation, and likely forced the system into secondary swapping or logging activity. The silver narrative therefore emphasizes a network-driven resource cascade.
\end{minipage}\hfill
\begin{minipage}[t]{0.48\linewidth}
\textbf{\frameworkname exported artifact.}\\
\small
Termination status: \texttt{resolved}.\\
Current best: disk write saturation caused high iowait and degraded responsiveness.\\
Retained alternatives: one weaker memory-centered explanation remains visible in the artifact.\\
Paper-facing value: a short case in which the engine reaches clean closure while still preserving a compact adjudication trace.
\end{minipage}
\end{tcolorbox}

\begin{tcolorbox}[statbox,title={Case 6\_2: appendix-facing stats summary}]
\small
\textbf{Session time:} 200.1s \quad
\textbf{Turns:} 2 \quad
\textbf{Tool calls:} 25 \quad
\textbf{LLM calls:} 12 \quad
\textbf{Tokens:} 63,277 \\
\textbf{Viewed metrics:} 20 / 53 (\(37.7\%\)) \quad
\textbf{Nodes:} 44 \quad
\textbf{Edges:} 48 \quad
\textbf{Debug signals:} 27 \\
\textbf{Alternatives retained:} 2 \quad
\textbf{Key conflicts:} 0 \quad
\textbf{Coverage ratio (best hypothesis):} 1.0
\end{tcolorbox}

\begin{tcolorbox}[eventbox,title={Case 6\_2: graph-evolution / event excerpt}]
\small
\textbf{Event 1: alternative payload proposed.} The initial orchestration step explicitly introduced two challenger explanations: a user-space CPU explanation and a swap-driven explanation.\\[0.35em]
\textbf{Event 2: expert claim assessment.} The storage expert returned repeated \texttt{supports} assessments for saturation, iowait dominance, and blocked-process correlation, quickly separating the storage hypothesis from its competitors.\\[0.35em]
\textbf{Event 3: latest termination assessment.} The final debug state marked the case as \texttt{resolved} because the best hypothesis was well supported and the alternatives were sufficiently adjudicated, with zero key conflicts and full coverage for the current-best hypothesis.
\end{tcolorbox}

\begin{figure}[p]
    \centering
    \includegraphics[width=\textwidth,height=0.92\textheight,keepaspectratio]{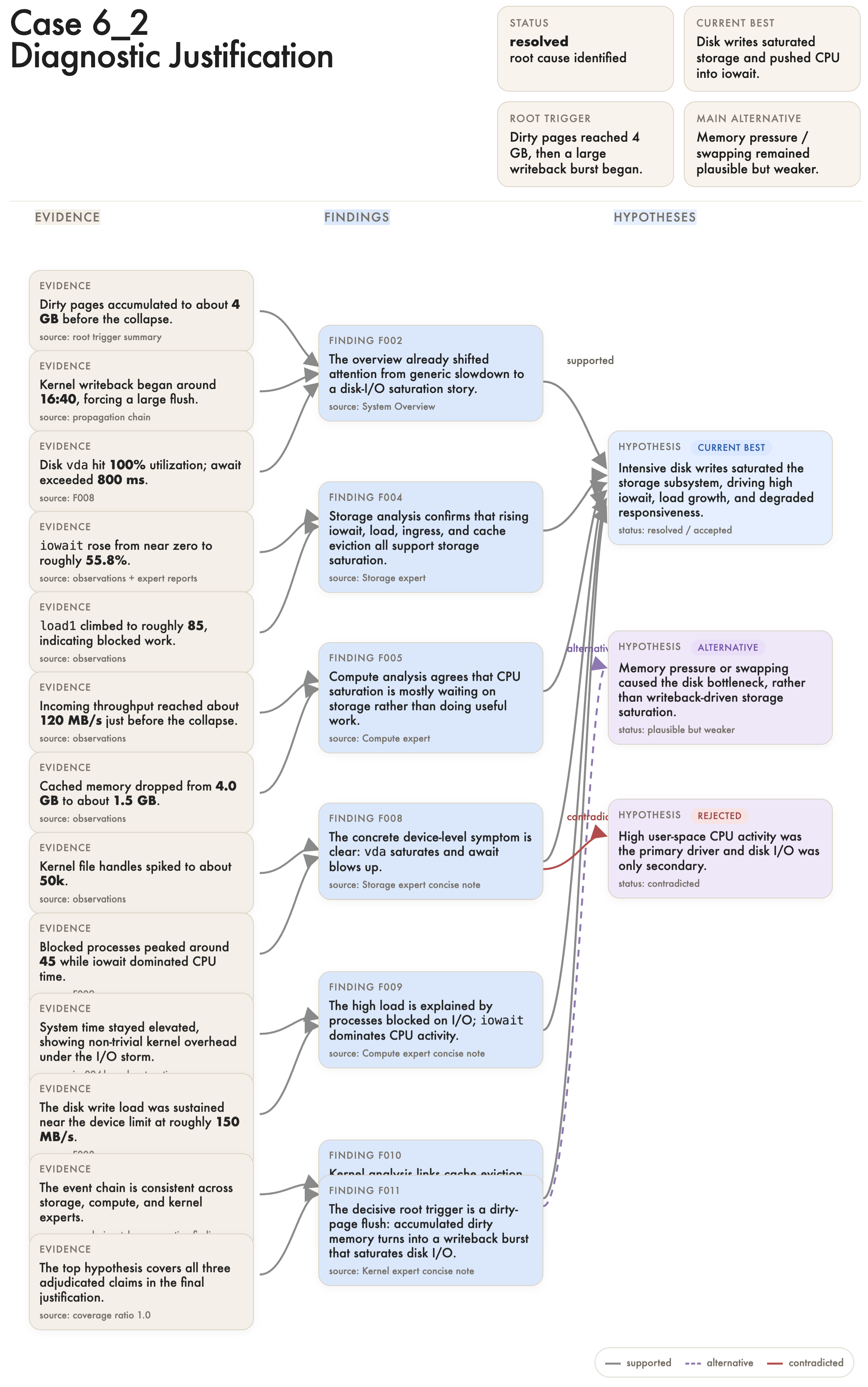}
    \caption{\textbf{Case 6\_2 full diagnostic justification artifact.}
    This resolved example complements Case~7 by showing the same export format in a cleaner setting. The artifact still preserves alternatives and claim-level structure, but the current-best storage explanation is sufficiently supported to justify \texttt{resolved} termination.}
    \label{fig:appendix-case6-2-a4}
\end{figure}

\subsection{Abridged Mechanism Prompts}

The boxes below are paper-facing excerpts from the manager-side prompt logic in \texttt{src/insightflow\_reborn/main.py}. We include them because they encode two of the paper's central design commitments: structured multi-hypothesis verification and adjudication-aware termination.

\begin{tcolorbox}[promptbox,title={Prompt excerpt A: full DJ control and structured verification}]
\ttfamily\small
DJ control mode is FULL: use claim-level adjudication, multi-hypothesis competition, \\
and explicit coverage-aware termination. \\
- Provide 2--4 total hypotheses (primary + alternatives), with count chosen by case complexity. \\
- In \texttt{propose\_expert\_verification}, \texttt{key\_claims\_to\_verify} MUST be claim objects \\
  and alternatives MUST use structured objects. \\
- In FULL mode, every hypothesis must be verified with equal scrutiny, not just the primary one.
\end{tcolorbox}

\begin{tcolorbox}[promptbox,title={Prompt excerpt B: adjudication-style termination contract}]
\ttfamily\small
2. After each investigation cycle, use \texttt{check\_termination\_conditions}. \\
3. If termination conditions are met for \texttt{resolved} or \texttt{provisionally\_resolved}, \\
   you MUST immediately use \texttt{propose\_termination}. \\
4. If the tool indicates \texttt{stalled}, do NOT present it as a clean success. \\
   Refresh or refine the hypotheses once if requested; otherwise terminate only with \\
   \texttt{insufficient\_information}. \\
5. Do not rely on free-text phrases to imply alternatives or contradictions; encode them in structured fields.
\end{tcolorbox}

\begin{tcolorbox}[promptbox,title={Prompt excerpt C: evidence-grounding and whiteboard discipline}]
\ttfamily\small 
**CRITICAL REQUIREMENTS:** \\
- Evidence Chain: All findings must be supported by structured evidence. \\
- Whiteboard Central: All diagnostic progress tracked on the diagnostic whiteboard. \\
- Use \texttt{update\_whiteboard(finding\_content="...", source="...", confidence=0.0)} \\
  to verify and add new findings to the shared diagnostic state.
\end{tcolorbox} 

\subsection{Priority Weighting in Claim Adjudication}

The main paper states that claims carry a priority field, but does not spell out the exact weighting coefficients. In the implementation, claim-level adjudication uses three priority tiers only: \texttt{high}, \texttt{medium}, and \texttt{low}. There is no separate \texttt{weak} priority tier; the word \texttt{weak} appears elsewhere only as a graph relation label.

For each structured claim assessment, the accumulated vote is:
\[
\mathrm{weighted\_vote}(a,c)
=
\mathrm{clip}_{[0,1]}(\mathrm{confidence}(a))
\cdot
w(\mathrm{priority}(c)),
\]
where the priority multiplier is
\[
w(\texttt{high}) = 1.5,\qquad
w(\texttt{medium}) = 1.0,\qquad
w(\texttt{low}) = 0.75.
\]
The adjudication store then sums weighted votes into the three verdict buckets \texttt{supports}, \texttt{contradicts}, and \texttt{insufficient}. Hypothesis-level scores are computed from these aggregated claim totals, so higher-priority claims contribute more strongly to the final ranking.

\subsection{Hypothesis Score Snapshots Across Four Cases}

To make the DJ adjudication process concrete, Table~\ref{tab:appendix-score-snapshots} summarizes four representative score snapshots from the full \texttt{main\_full} run. We intentionally choose a compact resolved case (\texttt{case6\_2}), a calibrated unresolved case (\texttt{case7}), a provisionally resolved case with live alternatives (\texttt{casesyshigh\_29}), and a resolved case with one plausible and one negative challenger (\texttt{case32\_1}).

\begin{table*}[t]
\centering
\caption{Four-case snapshots of hypothesis-level adjudication. Scores are the aggregated claim-level values recorded in the exported justification artifact.}
\label{tab:appendix-score-snapshots}
\scriptsize
\setlength{\tabcolsep}{4pt}
\begin{tabular}{@{}lcccc@{}}
\toprule
\textbf{Case} & \textbf{Termination} & \textbf{Top hypothesis} & \textbf{2nd hypothesis} & \textbf{3rd hypothesis} \\
\midrule
\texttt{case6\_2} & resolved & current\_best, \(18.73\) & plausible\_alternative, \(0.98\) & contradicted, \(-1.20\) \\
\texttt{case7} & stalled & current\_best, \(67.27\) & rejected, \(-5.85\) & rejected, \(-15.00\) \\
\texttt{casesyshigh\_29} & provisionally\_resolved & current\_best, \(44.31\) & plausible\_alternative, \(7.14\) & plausible\_alternative, \(3.85\) \\
\texttt{case32\_1} & resolved & current\_best, \(21.92\) & plausible\_alternative, \(2.55\) & unlikely, \(-0.85\) \\
\bottomrule
\end{tabular}
\end{table*}

\begin{tcolorbox}[statbox,title={Case 6\_2: why the storage hypothesis wins cleanly}]
\small
\textbf{Winning hypothesis.} The current-best storage hypothesis reached a score of \(18.73\), driven by three mutually reinforcing claims: disk saturation \(+\)6.90, dominant iowait \(+\)7.43, and write-driven pressure \(+\)4.40.\\[0.35em]
\textbf{Alternative behavior.} The swap-oriented challenger remained only weakly positive (\(0.98\)) because one claim stayed insufficient (\(1.05\)), while the user-space-CPU challenger became negative (\(-1.20\)) after contradiction against the observed iowait profile. This is a clean example of resolved closure with visible but weaker alternatives.
\end{tcolorbox}

\begin{tcolorbox}[statbox,title={Case 7: high support for the winner, but still not enough to close}]
\small
\textbf{Winning hypothesis.} The connection-storm hypothesis reached a very large positive score (\(67.27\)), supported by four repeated claims: temporal precedence of \texttt{ssEstab} (\(+23.85\)), kernel-dominated CPU time (\(+22.43\)), blocked-progress explanation (\(+12.35\) net), and crash / restart evidence (\(+11.55\)).\\[0.35em]
\textbf{Why the case still stalls.} The storage challenger fell to \(-15.00\) and the memory/OOM challenger to \(-5.85\), yet the final artifact still preserved them for reviewer visibility. A remaining conflict edge and one missing discriminative check prevented responsible clean closure, so the run terminated as \texttt{stalled} rather than converting a high score into premature certainty.
\end{tcolorbox}

\begin{tcolorbox}[statbox,title={Case \texttt{casesyshigh\_29}: provisionally resolved with two live challengers}]
\small
\textbf{Winning hypothesis.} The storage-centered explanation scored \(44.31\), driven mainly by the temporal lead of iowait before the load spike (\(+22.13\)) and the strong correlation between blocked processes and disk latency (\(+22.19\)).\\[0.35em]
\textbf{Why alternatives remain live.} Two challengers remained positive rather than being eliminated outright: a network-first explanation (\(7.14\)) and a weaker third explanation (\(3.85\)). The strongest unresolved weakness of the current-best hypothesis was its failed OOM-style memory claim, which contributed \(9.30\) contradiction. This is exactly the kind of score pattern that leads to \texttt{provisionally\_resolved} rather than fully resolved output.
\end{tcolorbox}

\begin{tcolorbox}[statbox,title={Case \texttt{case32\_1}: resolved with one plausible and one negative challenger}]
\small
\textbf{Winning hypothesis.} The current-best hypothesis scored \(21.92\), supported by a linear memory-growth claim (\(+5.85\)), a process-start / workload-shift claim (\(+9.84\)), and persistence of the post-event memory state (\(+2.89\)), together with an additional supportive background-task separation claim.\\[0.35em]
\textbf{Challenger behavior.} One alternative remained plausible at \(2.55\), but was materially weakened by contradiction on traffic direction (\(3.95\) contradiction on the outbound/inbound claim). A second challenger became negative (\(-0.85\)) because its kernel-level leak explanation was contradicted by the observed stable buffer/cache behavior. This case is useful because it shows how DJ can still produce a resolved output while making the score gaps among alternatives explicit.
\end{tcolorbox}

\subsection{Dataset-Level Graph Statistics}

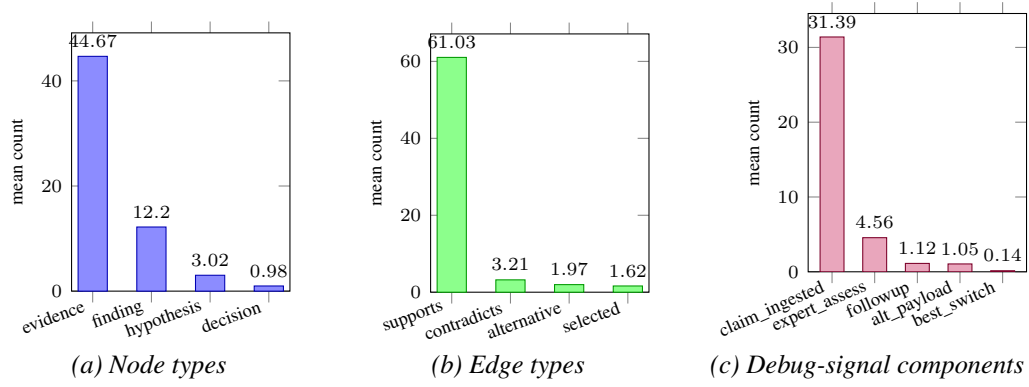
\begin{figure*}[t]
\centering
\begin{minipage}[t]{0.32\textwidth}
\centering
\begin{tikzpicture}
\begin{axis}[
    ybar,
    bar width=11pt,
    width=\linewidth,
    height=5.0cm,
    ymin=0,
    ylabel={mean count},
    symbolic x coords={evidence,finding,hypothesis,decision},
    xtick=data,
    xticklabel style={font=\scriptsize, rotate=20, anchor=east},
    yticklabel style={font=\scriptsize},
    label style={font=\scriptsize},
    nodes near coords,
    every node near coord/.append style={font=\scriptsize},
    enlarge x limits=0.12
]
\addplot[fill=blue!45!white, draw=blue!70!black] coordinates {
    (evidence,44.67)
    (finding,12.20)
    (hypothesis,3.02)
    (decision,0.98)
};
\end{axis}
\end{tikzpicture}

\vspace{-0.3em}
\textit{(a) Node types}
\end{minipage}\hfill
\begin{minipage}[t]{0.32\textwidth}
\centering
\begin{tikzpicture}
\begin{axis}[
    ybar,
    bar width=11pt,
    width=\linewidth,
    height=5.0cm,
    ymin=0,
    ylabel={mean count},
    symbolic x coords={supports,contradicts,alternative,selected},
    xtick=data,
    xticklabel style={font=\scriptsize, rotate=20, anchor=east},
    yticklabel style={font=\scriptsize},
    label style={font=\scriptsize},
    nodes near coords,
    every node near coord/.append style={font=\scriptsize},
    enlarge x limits=0.12
]
\addplot[fill=green!45!white, draw=green!60!black] coordinates {
    (supports,61.03)
    (contradicts,3.21)
    (alternative,1.97)
    (selected,1.62)
};
\end{axis}
\end{tikzpicture}

\vspace{-0.3em}
\textit{(b) Edge types}
\end{minipage}\hfill
\begin{minipage}[t]{0.32\textwidth}
\centering
\begin{tikzpicture}
\begin{axis}[
    ybar,
    bar width=9pt,
    width=\linewidth,
    height=5.0cm,
    ymin=0,
    ylabel={mean count},
    symbolic x coords={claim\_ingested,expert\_assess,followup,alt\_payload,best\_switch},
    xtick=data,
    xticklabel style={font=\scriptsize, rotate=24, anchor=east},
    yticklabel style={font=\scriptsize},
    label style={font=\scriptsize},
    nodes near coords,
    every node near coord/.append style={font=\scriptsize},
    enlarge x limits=0.14
]
\addplot[fill=purple!35!white, draw=purple!65!black] coordinates {
    (claim\_ingested,31.39)
    (expert\_assess,4.56)
    (followup,1.12)
    (alt\_payload,1.05)
    (best\_switch,0.14)
};
\end{axis}
\end{tikzpicture}

\vspace{-0.3em}
\textit{(c) Debug-signal components}
\end{minipage}
\caption{\textbf{Dataset-level artifact structure for the full \texttt{main\_full} run.}
Panel (a) shows the average node composition per case, Panel (b) shows the average edge composition, and Panel (c) shows the most frequent debug-signal components extracted from \texttt{diagnostic\_graph\_debug.json}. The distributions reflect a strongly evidence-heavy graph, a predominantly supportive edge structure with localized contradiction retention, and repeated claim-assessment updates as the dominant event-level activity.}
\label{fig:appendix-graph-stats}
\end{figure*}

To complement the case-level appendix views, we summarize the structure of the exported diagnostic justification artifacts over the full \texttt{main\_full} run. Across 66 cases, the system produced on average \(60.9 \pm 18.5\) nodes, \(67.8 \pm 21.1\) edges, and \(38.3 \pm 13.2\) debug signals per case. Figure~\ref{fig:appendix-graph-stats} breaks these totals down by node type, edge type, and the most common debug-signal components.

\subsection{How These Artifact Examples Support the Main Paper}

These appendix cases play a narrow but important role. Case~7 supports the main paper's argument that calibrated non-closure can still yield a useful, reviewable diagnosis artifact. Case~6\_2 shows the complementary mode in which the same machinery produces a compact resolved artifact rather than a stalled one. The prompt excerpts then connect these visible outputs back to the mechanism design in the main text: explicit expert dispatch, claim review, retained alternatives, and adjudication-aware termination.

\newpage
\section{Judge Protocol and Prompt Details}
\label{sec:appendix-judge}

Our evaluation protocol uses two complementary judges. The outcome judge evaluates only the exported final diagnosis artifact, whereas the process judge evaluates the structured justification artifact. This separation is intentional: a plausible final answer does not by itself establish that the reasoning process was accountable, evidence-linked, or uncertainty-aware.

\subsection{Silver Reference Narrative}

The reference used in judging is a \emph{silver reference narrative}, not a gold ground-truth label. In our setting, many incidents do not admit a perfectly objective, fully observed root-cause statement. The reference therefore serves as a strong semantic anchor derived from real SRE-style incident analysis, while the prompts explicitly instruct the judge not to treat it as infallible ground truth. This allows the judge to give bounded partial credit when the final diagnosis diverges from the reference narrative but preserves relevant alternatives, contradictions, or uncertainty with meaningful evidence support.

\subsection{Outcome-Only Judge}

The outcome judge reads the final exported fields from \texttt{diagnosis\_conclusion.json}, including the fault summary, propagation chain, root triggers, key metrics, root-cause analysis, recommendations, and termination status. It also receives a small amount of structured process context, such as the current best hypothesis, plausible alternatives, key conflicts, missing distinguishing evidence, remaining uncertainties, and a compact claim-adjudication summary. This additional context allows the judge to recognize when a system preserves evidence-backed alternatives without fully collapsing them into the final narrative.

The outcome judge scores four dimensions: correctness, evidence grounding, causal coherence, and actionability. The composite outcome score is a weighted arithmetic mean:
\[
\text{OutcomeScore} = 0.40 \cdot \text{Correctness} + 0.25 \cdot \text{EvidenceGrounding} + \\ 0.25 \cdot \text{CausalCoherence} + 0.10 \cdot \text{Actionability}.
\]

\subsection{Process-Aware Judge}

The process judge reads a compressed structured artifact rather than the raw conversation trace. Its inputs include the termination status, current best hypothesis, plausible alternatives, key supporting findings, key conflicts, missing distinguishing evidence, remaining uncertainties, recommended next checks, and aggregate process metrics from the compat export. This judge is designed to score whether the system maintained an accountable diagnostic process, not merely whether it produced a plausible final story.

The process judge evaluates five dimensions: evidence grounding, alternative handling, contradiction handling, traceability, and uncertainty management. The composite process score is the simple arithmetic mean of these five dimensions. We intentionally do not use a harmonic mean, because these dimensions represent complementary evaluation facets rather than tightly coupled rates such as precision and recall.

\subsection{What the Judge Sees and Does Not See}

The judges operate on exported artifacts, not on hidden chain-of-thought text. This design improves reproducibility and reduces dependence on brittle prompt traces. It also makes clear what is and is not being evaluated:

\begin{itemize}
    \item The outcome judge sees the final diagnosis and limited structured context about alternatives and conflicts.
    \item The process judge sees the compressed justification artifact and compat process metrics.
    \item Neither judge sees the full raw multi-agent chat log by default.
\end{itemize}

Accordingly, low process scores for some baselines should be interpreted as weakness in explicit accountable artifact support, not necessarily as proof that the method could never produce a plausible narrative answer.

\newpage
\section{Expanded Experimental Results}
\label{sec:appendix-results} 

This section reports the full mean \(\pm\) standard deviation values used to summarize the main, ablation, and sensitivity studies. Unless otherwise noted, the standard deviations are computed across cases within each setting, not across repeated runs. We omit backbone mini checks and turn-budget sweeps here to keep the appendix focused on the main paper's central comparisons.

\subsection{Main Results with Case-Level Variability}

\begin{table}[h]
\centering
\caption{Main full-dataset results with case-level standard deviations.}
\label{tab:appendix-main-std}
\scriptsize
\setlength{\tabcolsep}{4pt} 
\begin{tabular}{@{}lcccccc@{}}
\toprule
\textbf{Method} & \textbf{Outcome} & \textbf{Process} & \textbf{Time (s)} & \textbf{Turns} & \textbf{Tool Calls} & \textbf{Tokens (K)} \\
\midrule
\rcagent & 44.3\styledstd{24.5} & 9.5\styledstd{4.1} & 64.3\styledstd{58.1} & 2.7\styledstd{1.4} & 6.6\styledstd{2.4} & 21.7\styledstd{21.9} \\
\flowofaction & 42.8\styledstd{21.7} & 9.3\styledstd{5.6} & 276.3\styledstd{172.2} & 4.0\styledstd{4.3} & 19.7\styledstd{9.6} & 92.9\styledstd{94.2} \\
\texttt{Main w/o DJ} & 51.0\styledstd{21.3} & 44.0\styledstd{13.5} & 315.7\styledstd{127.5} & 5.8\styledstd{2.9} & 23.5\styledstd{7.8} & 154.3\styledstd{82.8} \\
\frameworkname & \textbf{57.7}\styledstd{20.4} & \textbf{50.5}\styledstd{12.8} & 457.0\styledstd{177.7} & 6.5\styledstd{3.8} & 33.7\styledstd{11.3} & 218.3\styledstd{130.7} \\
\bottomrule
\end{tabular}
\end{table}

\subsection{Mini-Subset Ablations with Case-Level Variability}

\begin{table}[h]
\centering
\caption{Ablation results on the 10-case mini subset with case-level standard deviations.}
\label{tab:appendix-ablation-std}
\scriptsize
\setlength{\tabcolsep}{4pt}
\begin{tabular}{@{}lcccccc@{}}
\toprule
\textbf{Setting} & \textbf{Outcome} & \textbf{Process} & \textbf{Time (s)} & \textbf{Turns} & \textbf{Tool Calls} & \textbf{Tokens (K)} \\
\midrule
\frameworkname~(mini) & 53.6\styledstd{22.2} & 53.8\styledstd{15.0} & 472.4\styledstd{172.1} & 6.9\styledstd{3.9} & 33.8\styledstd{9.4} & 188.9\styledstd{86.1} \\
\texttt{w/o claim adjudication} & 45.9\styledstd{19.0} & 45.6\styledstd{12.9} & 364.1\styledstd{138.9} & 6.7\styledstd{3.6} & 25.5\styledstd{8.1} & 171.2\styledstd{80.6} \\
\texttt{w/o evidence grounding} & 49.1\styledstd{23.4} & 35.4\styledstd{15.1} & 555.5\styledstd{136.7} & 5.9\styledstd{4.5} & 40.2\styledstd{10.1} & 233.9\styledstd{115.8} \\
\texttt{w/o hypothesis competition} & \textbf{55.8}\styledstd{18.3} & 50.4\styledstd{12.3} & 307.3\styledstd{150.2} & 5.4\styledstd{1.8} & 24.7\styledstd{12.8} & 162.1\styledstd{82.1} \\
\texttt{w/o coverage gate} & 54.1\styledstd{22.0} & 50.1\styledstd{11.3} & 348.7\styledstd{198.0} & 4.9\styledstd{1.7} & 21.3\styledstd{5.8} & 154.1\styledstd{66.4} \\
\bottomrule
\end{tabular}
\end{table}

\subsection{Sensitivity Results with Case-Level Variability}

\begin{table}[h]
\centering
\caption{Sensitivity results on the 10-case mini subset with case-level standard deviations.}
\label{tab:appendix-sensitivity-std}
\scriptsize
\setlength{\tabcolsep}{4pt}
\begin{tabular}{@{}lcccccc@{}}
\toprule
\textbf{Setting} & \textbf{Outcome} & \textbf{Process} & \textbf{Time (s)} & \textbf{Turns} & \textbf{Tool Calls} & \textbf{Tokens (K)} \\
\midrule
\(C=2\) & 52.0\styledstd{19.3} & 43.2\styledstd{12.0} & 344.3\styledstd{110.6} & 5.4\styledstd{2.7} & 21.9\styledstd{6.5} & 176.1\styledstd{81.9} \\
\(C=3\) & 60.8\styledstd{16.4} & 51.2\styledstd{16.2} & 424.9\styledstd{113.3} & 5.9\styledstd{3.4} & 27.8\styledstd{8.0} & 193.4\styledstd{62.0} \\
\(C=4\) & 60.8\styledstd{16.1} & \textbf{60.6}\styledstd{11.5} & 693.0\styledstd{272.9} & 8.5\styledstd{4.4} & 39.9\styledstd{17.2} & 256.9\styledstd{131.7} \\
\(K=2\) & \textbf{63.6}\styledstd{24.3} & 53.1\styledstd{13.8} & 449.0\styledstd{185.0} & 7.7\styledstd{4.1} & 32.1\styledstd{13.9} & 213.9\styledstd{84.3} \\
\(K=3\) & 55.7\styledstd{20.5} & 53.9\styledstd{18.6} & 513.5\styledstd{244.1} & 8.3\styledstd{3.9} & 37.9\styledstd{18.5} & 249.2\styledstd{96.0} \\
\(K=4\) & 53.2\styledstd{25.9} & 60.0\styledstd{12.8} & 1213.4\styledstd{595.1} & 11.3\styledstd{4.5} & 56.1\styledstd{30.1} & 340.2\styledstd{168.0} \\
\(K=2\text{--}4\) & 59.8\styledstd{23.6} & 54.1\styledstd{10.9} & 615.5\styledstd{306.6} & 6.9\styledstd{3.9} & 36.8\styledstd{16.9} & 259.4\styledstd{277.5} \\
\(\tau=0.5\) & 59.7\styledstd{28.2} & \textbf{55.4}\styledstd{13.5} & 421.1\styledstd{201.3} & 7.3\styledstd{3.5} & 29.0\styledstd{13.7} & 210.3\styledstd{108.9} \\
\(\tau=0.67\) & 59.5\styledstd{22.7} & 48.2\styledstd{12.8} & 411.2\styledstd{170.1} & 6.4\styledstd{3.6} & 27.6\styledstd{11.2} & 192.5\styledstd{117.9} \\
\(\tau=0.8\) & \textbf{63.9}\styledstd{18.1} & 48.4\styledstd{14.2} & 641.6\styledstd{252.5} & 6.7\styledstd{2.9} & 33.1\styledstd{11.1} & 210.5\styledstd{62.7} \\
\bottomrule
\end{tabular}
\end{table}



\end{document}